\documentclass[aps,onecolumn,groupedaddress,12pt]{revtex4-1}
\usepackage{amsmath}
\usepackage{amssymb}
\usepackage{graphicx}
\usepackage{dcolumn}
\usepackage{bm}

\begin{document}

\title{Bound states in the continuum accompanied by avoided crossings in leaky-mode photonic lattices}

\author{Sun-Goo Lee}
\email{sungooleee@gmail.com}
\author{Seong-Han Kim}
\author{Chul-Sik Kee}
\email{cskee@gist.ac.kr}
\affiliation{Integrated Optics Laboratory, Advanced Photonics Research Institute, GIST, Gwangju 61005, South Korea}
\date{\today}

\begin{abstract}
When two nonorthogonal resonances are coupled to the same radiation channel, avoided crossing arises and a bound state in the continuum (BIC) appears in parametric space. This paper presents numerical and analytical results on the properties of avoided crossing and BIC due to the coupled guided-mode resonances in one-dimensional leaky-mode photonic lattices with slab geometry. In symmetric photonic lattices with up-down mirror symmetry, Friedrich-Wintgen BICs with infinite lifetime are accompanied by avoided crossings due to the coupling between two guided modes with the same transverse parity. In asymmetric photonic lattices with broken up-down mirror symmetry, quasi-BICs with finite lifetime appear with avoided crossings because radiating waves from different modes cannot be completely eliminated. We also show that unidirectional-BICs are accompanied by avoided crossings due to guided-mode resonances with different transverse parities in asymmetric photonic lattices. The $Q$ factor of a unidirectional-BIC is finite, but its radiation power in the upward or downward direction is significantly smaller than that in the opposite direction. Our results may be helpful in engineering BICs and avoided crossings in diverse photonic systems that support leaky modes.
\end{abstract}

%\pacs{78.67-n, 42.70.Qs}

\maketitle

\section{Introduction}
The ability to confine light to limited regions is of fundamental importance in both basic science and practical applications. Conventionally, electromagnetic waves can be localized in photonic structures by separating specific eigenmodes away from the continuum of radiating modes. This mode separation can typically be achieved through metallic mirrors, total internal reflections at dielectric interfaces \cite{Born2002}, and photonic band gaps in periodic structures \cite{Yablonovitch1987,Joannopoulos1995}. Optical bound states in the continuum (BICs) are special electromagnetic states that remain well localized in photonic structures even though they coexist with outgoing waves that can carry electromagnetic energy away from the photonic structure \cite{Marinica2008,Plotnik2011,Hsu2016,Koshelev2019,Koshelev2020}. Diverse types of BICs have been implemented in various photonic systems, including metasurfaces \cite{Koshelev2018,Kupriianov2019,Abujetas2019}, photonic crystals \cite{Gansch2016,Yang2014}, plasmonic structures \cite{Azzam2018}, and fiber Bragg gratings \cite{XGao2019}. Recently, robust BICs in subwavelength photonic crystal slab geometry have attracted much attention because they are associated with interesting topological physical phenomena \cite{BZhen2014,YXGao2017,SGLee2019-1} as well as practical applications, such as lasers \cite{Kodigala2017,STHa2018}, sensors \cite{YLiu2017,Romano2018}, and filters \cite{Foley2014}.

BICs found in slab-type photonic lattices so far can be split into three categories: (i) symmetry-protected BICs, (ii) single-resonance parametric BICs, and  (iii) Friedrich-Wintgen BICs. Symmetry-protected BICs appear at the $\Gamma$ point (the center of the Brillouin zone) due to the symmetry mismatch between their mode profiles and those of external plane waves \cite{Kazarinov1985,SLi2019}. Single-resonance parametric BICs are found at generic $k$ points along dispersion curves when the relevant coupling to the radiation continuum completely vanishs \cite{Hsu2013}. Friedrich-Wintgen BICs, which are generally found in the vicinity of the avoided crossing of two dispersion curves, arise because of the destructive interference of two guided-mode resonances coupled to the same radiation channel \cite{Bulgakov2018}. Historically, Friedrich and Wintgen presented a general formalism to find BICs in quantum systems in 1985 \cite{Friedrich1985}. Recently, it has been shown that the Friedrich-Wintgen formalism is valid to describe optical BICs in photonic structures \cite{Mermet-Lyaudoz2019,Kikkawa2019}. The aim of the present paper is to address the fundamental properties of avoided crossings and BICs due to coupled guided-mode resonances in one-dimensional (1D) leaky-mode photonic lattices.

When two nonorthogonal resonances generate avoided crossings, BICs with infinite lifetimes appear in parametric space, and the conditions for Friedrich-Wintgen BICs can generally be fulfilled through the fine tuning of structural parameters. In photonic lattice slabs, however, the Friedrich-Wintgen BIC can be found near the avoided crossing in the photonic band structure without the fine tuning of structural parameters. In this study, we investigated BICs and avoided crossing due to two different waveguide modes in photonic lattice slabs with symmetric and asymmetric cladding layers through finite element method (FEM) simulations and temporal coupled-mode formalism. We show that avoided crossings in the photonic lattices with asymmetric cladding layers support only quasi-BICs with a finite value of $Q$ factor, whereas the avoided crossings with symmetric cladding structures support true-BICs with infinite $Q$ factor. We also show that unidirectional-BICs are accompanied by avoided crossings due to two guided-mode resonances with different transverse parities in asymmetric photonic lattices. The $Q$ factor of the unidirectional-BIC is finite but its radiation power in the upward or downward direction is significantly smaller than that in the opposite direction.

\section{Lattice structure and perspective}
\begin{figure}[]
\centering\includegraphics[width=14cm]{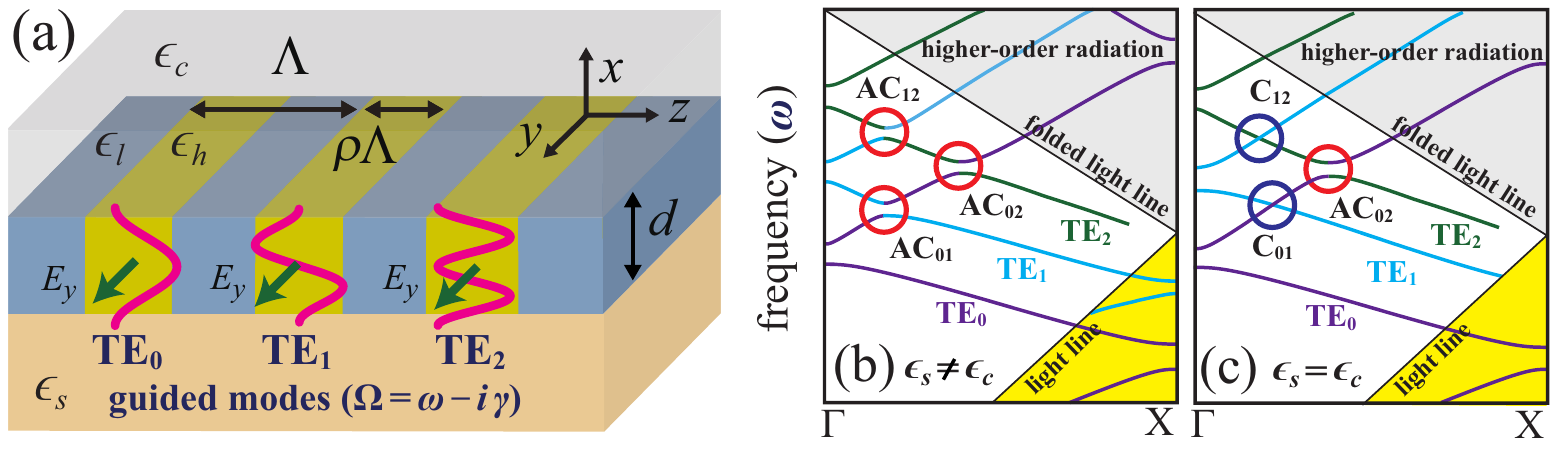}
\caption{(a) Schematic of a 1D photonic lattice for studying avoided crossings and bound states. With periodic dielectric constant modulation, guided modes are described by the complex frequency $\Omega = \omega - i \gamma$, where $\gamma$ represents the decay rate of the mode. Conceptual illustration of the photonic band structures including avoided crossings due to different waveguide modes in the photonic lattices (b) with asymmetric cladding layers ($\epsilon_{s}\neq\epsilon_{c}$) and (c) symmetric cladding layers ($\epsilon_{s}= \epsilon_{c}$).}
\label{fig1}
\end{figure}

Figure~\ref{fig1} illustrates a 1D photonic lattice and the attendant schematic photonic band structures including avoided crossings. As shown in Fig.~\ref{fig1}(a), we model a 1D photonic lattice consisting of high ($\epsilon_{h}$) and low ($\epsilon_{l}$) dielectric constant media. A single periodic layer of thickness $d$ is enclosed by a substrate medium (lower cladding) of dielectric constant $\epsilon_{s}$ and cover (upper cladding) of $\epsilon_{c}$. The period of the lattice is $\Lambda$ and width of high dielectric constant medium is $\rho \Lambda$. This simple lattice supports multiple TE-polarized guided modes, and each mode has its own dispersion curve because the thickness $d=1.30~\Lambda$ is thick enough and its average dielectric constant $\epsilon_{avg}=\epsilon_{l}+\rho (\epsilon_{h}- \epsilon_{l})=6.00$ is larger than $\epsilon_{s}$ and $\epsilon_{c}$ \cite{Magnusson2009}. In dielectric slab waveguides with symmetric (asymmetric) cladding layers $\epsilon_{s}=\epsilon_{c}$ ($\epsilon_{s}\neq \epsilon_{c}$), as schematically illustrated in Fig.~\ref{fig1}(a), guided modes are classified into two categories by their transverse mode profiles \cite{Agrawal2004}. Even (even-like) modes $\mathrm{TE}_{m=0,2,4\cdots}$ have even (even-like) transverse electric field profiles, and odd (odd-like) modes $\mathrm{TE}_{m=1,3,5\cdots}$ have odd (old-like) transverse field profiles with symmetric (asymmetric) cladding layers. In  photonic lattices with asymmetric cladding layers, as shown in Fig.~\ref{fig1}(b), avoided crossings $\mathrm{AC}_{mn}$ (in red circles) due to $\mathrm{TE}_{m}$ and $\mathrm{TE}_{n}$ modes arise when $0<\rho<1$ and $\Delta \epsilon=\epsilon_{h} - \epsilon_{l}> 0$. In photonic lattices with symmetric cladding layers, as shown in Fig.~\ref{fig1}(c), two even modes generate avoided crossing $\mathrm{AC}_{02}$ (in red circle), but dispersion curves due to even and odd modes cross each other ($\mathrm{C}_{01}$ and $\mathrm{C}_{02}$ in blue circles) because even and odd modes are perfectly orthogonal in symmetric waveguide structures. In this study, we limited our attention to the avoided crossings $\mathrm{AC}_{01}$ and $\mathrm{AC}_{02}$ in asymmetric photonic lattices ($\epsilon_{s}=2.25$ and $\epsilon_{c}=1.00$) and $\mathrm{AC}_{02}$ in symmetric lattices ($\epsilon_{s}=\epsilon_{c}=2.25$) because these simplest cases clearly demonstrate the key properties of the avoided crossings and BICs in photonic lattice slabs. We consider the avoided crossings only in the white region where quasi-guided modes can couple to external plane waves effectively and generate diverse zero-order spectral responses \cite{Niraula2015,YHKo2018,SGLee2017}. In the yellow region below the light line in the substrate, guided modes are nonleaky and not associated with BICs \cite{Johnson1999}. In the gray region above the folded light line, guided modes are less practical because they generate higher-order diffracted waves outside the lattice \cite{YDing2007}.

\section{Results and discussion}
\begin{figure}[]
\centering\includegraphics[width=16.0cm]{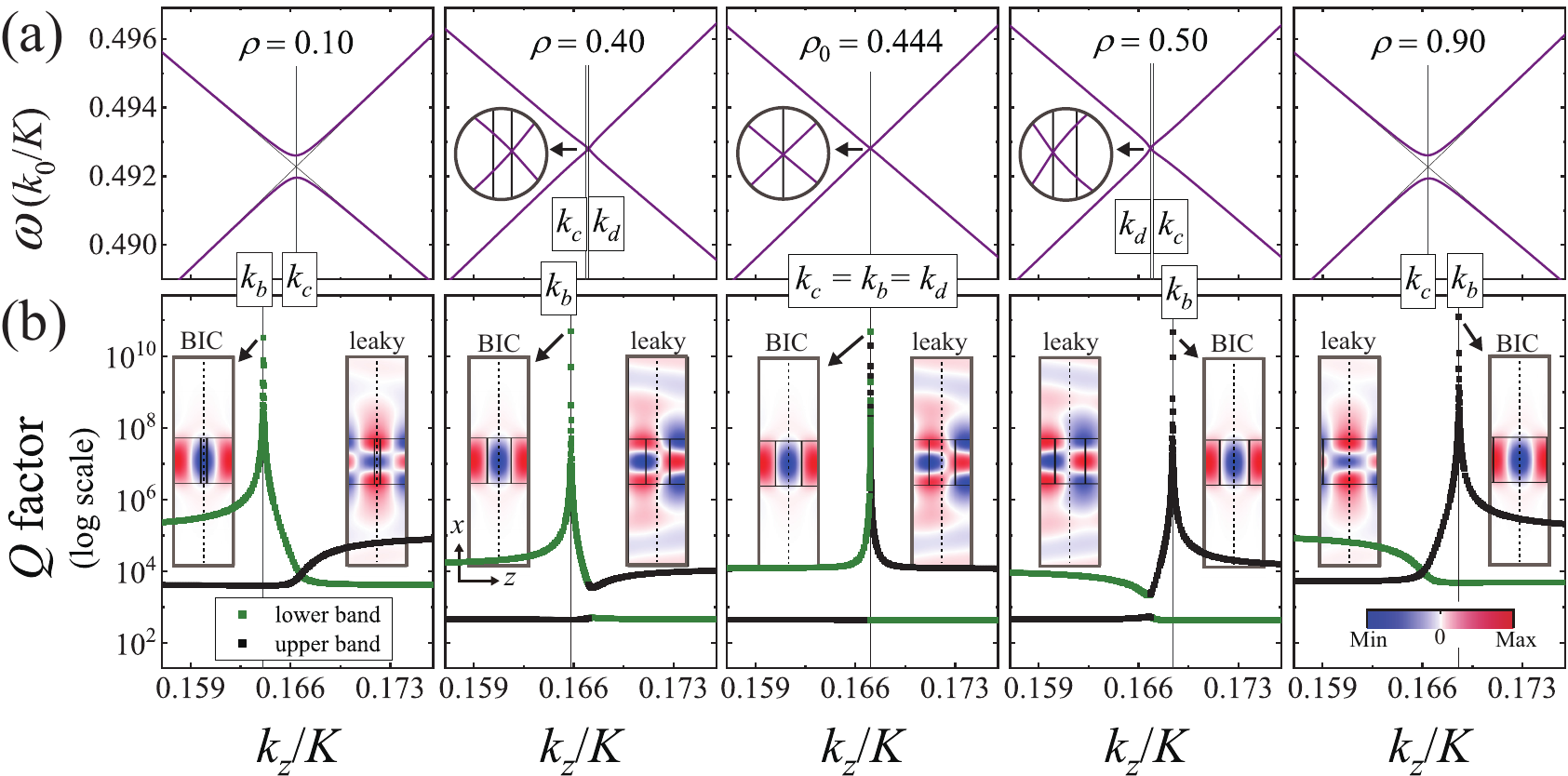}
\caption{Avoided crossings and BICs due to $\mathrm{TE}_{0}$ and $\mathrm{TE}_{2}$ modes in leaky-mode photonic lattices with symmetric cladding layers. (a) FEM simulated dispersion relations near avoided crossings for five different values of $\rho$.  Here, $k_{0}$ denotes the wavenumber in free space and $K=2\pi /\Lambda$ is the magnitude of the grating vector. Insets illustrate magnified views of dispersion curves near the crossing points. (b) Simulated $Q$ factors of guided modes in upper and lower bands. Insets with blue and red colors represent spatial electric field ($E_{y}$) distributions of BICs and leaky modes at the $y=0$ plane. Vertical dotted lines denote the mirror plane in the computational cell. In the FEM analysis, we use structural parameters $\epsilon_{avg}=6.00, \Delta \epsilon=1.00, d=1.30~\Lambda$, and $\epsilon_{s}=\epsilon_{c}=2.25$. }
\label{fig2}
\end{figure}

Figure~\ref{fig2}(a) shows the evolution of the avoided crossing $\mathrm{AC}_{02}$ due to $\mathrm{TE}_{0}$ and $\mathrm{TE}_{2}$ modes under variation of $\rho$ in the photonic lattice with symmetric cladding layers. As seen in Fig.~\ref{fig2}(a), a band gap opens at $k_c$ where two uncoupled dispersion curves cross each other, and its size increases as the value of $\rho$ increases from zero. However, the gap size decreases and becomes zero as $\rho$ is further increased. The bands remain closed for a while in spite of the additional increase in $\rho$. The band gap reopens and its size grows again, decreases, and approaches zero when $\rho$ is further increased and approaches 1. The Insets of Fig.~\ref{fig2}(a) depicting magnified views of the dispersion curves near the crossing point $k_c$ indicate that the degenerate point $k_d$ where the band closes is slightly different from $k_c$ in general. As $\rho$ increases, the relative position of $k_d$ changes from the right to left side of $k_c$. These band dynamics are associated with the band transition of the Friedrich-Wintgen BIC, as seen by the simulated $Q$ factors plotted in Fig.~\ref{fig2}(b). As $\rho$ increases from zero, the Friedrich-Wintgen BICs with $Q$ factors larger than $10^{10}$ appear at $k_b$ near the crossing point $k_c$. The distance between the location of the BIC and crossing point $|k_c-k_b|$ increases, decreases, and becomes zero when $\rho=0.444$. However, the distance increases again, decreases, and approaches zero as $\rho$ is further increased and approaches 1. The Friedrich-Wintgen BIC across the band gap under the variation of $\rho$ by passing through the degenerate point $k_b=k_c =k_d$ where two dispersion curves cross as straight lines. The spatial electric field ($E_y$) distributions plotted in the insets of  Fig.~\ref{fig2}(b) show that the Friedrich-Wintgen BICs, that have $\mathrm{TE}_{0}$-like field distributions, are well localized in the lattice without radiative loss, whereas leaky modes in the opposite band branch with $\mathrm{TE}_{2}$-like field distributions are radiative outside the lattice.

\begin{figure}[]
\centering\includegraphics[width=16.0cm]{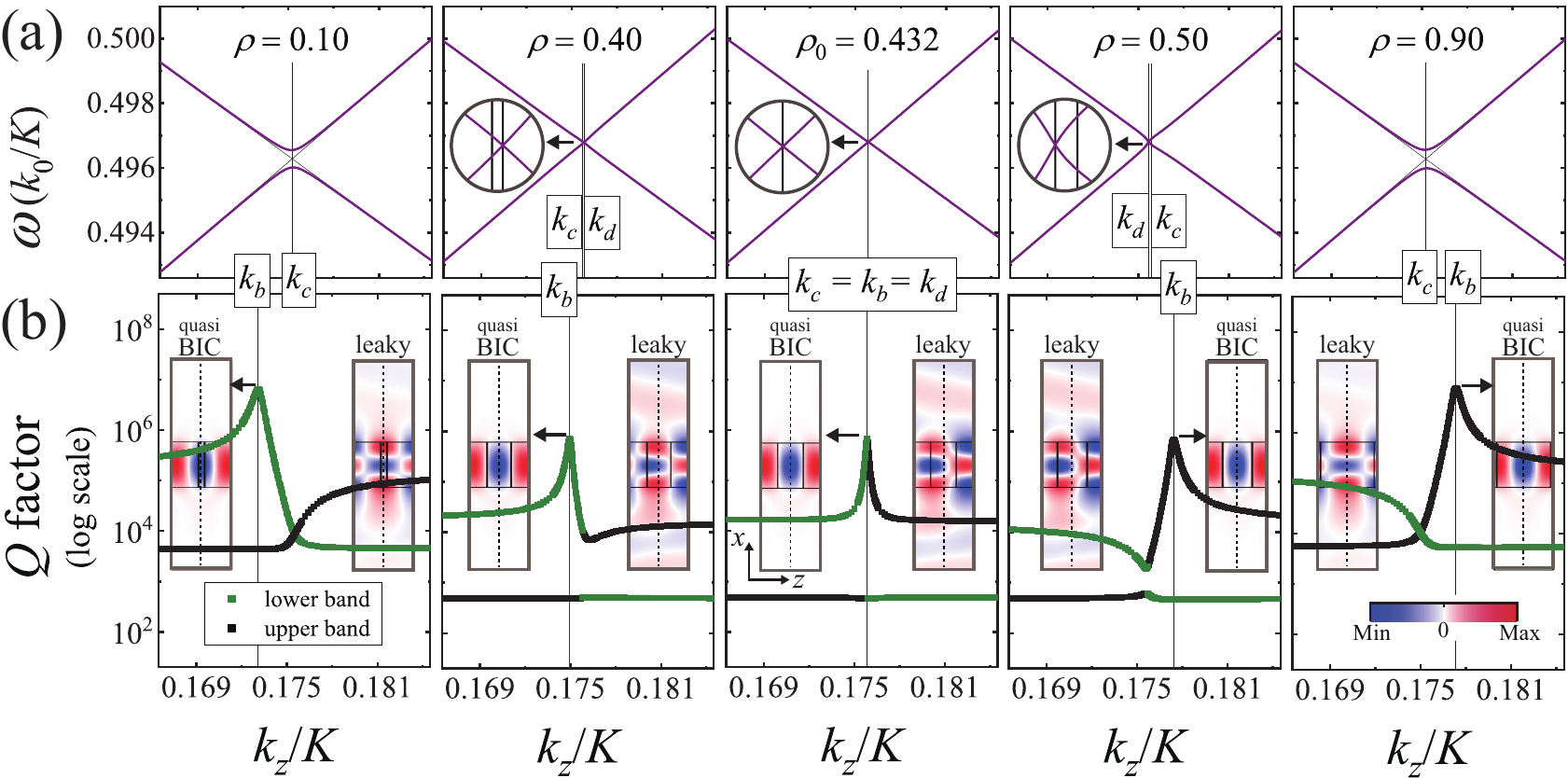}
\caption{Avoided crossings and quasi-BICs due to $\mathrm{TE}_{0}$ and $\mathrm{TE}_{2}$ modes in leaky-mode photonic lattices with asymmetric cladding layers. (a) Simulated dispersion relations near avoided crossings for five different values of $\rho$. Insets illustrate magnified views of dispersion curves near the crossing points. (b) Simulated $Q$ factors of guided modes in upper and lower bands. Insets with blue and red colors represent spatial electric field ($E_{y}$) distributions of BICs at the $y=0$ plane. Structural parameters are the same as in Fig.~\ref{fig2} except that $\epsilon_{c}=1.00$. }
\label{fig3}
\end{figure}

Figure~\ref{fig3}(a) illustrates the evolution of the avoided crossing $\mathrm{AC}_{02}$ due to $\mathrm{TE}_{0}$ and $\mathrm{TE}_{2}$ modes of photonic lattices with asymmetric cladding layers. The band dynamics shown in Fig.~\ref{fig3}(a) is the same as that in Fig.~\ref{fig2}(a). As $\rho$ varies from zero to 1, the band gap opens at $k_c$, closes at $k_d$, reopens, and vanishes with $\rho=1$. In the evolution process under variation of $\rho$, there exists a finite range of $\rho$ in which the bands remain closed. The degenerate point $k_d$ becomes the same as $k_c$ when the two dispersion curves cross as straight lines. In the closed band states with $k_c \neq k_d$, two dispersion curves have low curvatures, as clearly seen in the insets of Figs.~\ref{fig2}(a) and ~\ref{fig3}(a). The most noticeable effect of asymmetric cladding layers on the avoided crossings can be found by comparing the simulated $Q$ factors illustrated in Fig.~\ref{fig3}(b) with those in Fig.~\ref{fig2}(b). There exist quasi-BICs with $\mathrm{TE}_{0}$-like spatial electric field distributions around the crossing point $k_c$ in Fig.~\ref{fig3}(b). The $Q$ factors of the quasi-BICs in Fig.~\ref{fig3}(b) are saturated to finite values less than $10^7$ at $k_b$, whereas the $Q$ values of the Friedrich-Wintgen BICs in Fig.~\ref{fig2}(b) seem to diverge to infinity at $k_b$. The quasi-BICs also pass through the degenerate point $k_b=k_c =k_d$ and across the band gap under variation of $\rho$, as do the Friedrich-Wintgen BICs.

The dynamics of avoided crossing and the band transition of the bound states illustrated in Figs.~\ref{fig2} and \ref{fig3} can be understood from the temporal coupled-mode theory describing the interference of two different resonances in the same resonator \cite{WSuh2004}. When two leaky waveguide modes $\mathrm{TE}_{m}$ and $\mathrm{TE}_{n}$ with complex frequencies $\Omega_{m}=\omega_{m}-i\gamma_{m}$ and $\Omega_{n}=\omega_{n}-i\gamma_{n}$, respectively, are excited in the photonic lattice shown in Fig.~\ref{fig1}(a) by the incoming waves $|s_+\rangle$, two resonance amplitudes $\mathbf{A}=(A_{m}, A_{n})^{\rm{T}} $ evolve in time as $d\mathbf{A}/dt=-i \mathcal{H} \mathbf{A} + \mathcal{D}^T|s_+\rangle$ with the Hamiltonian $\mathcal{H}$ and coupling matrix $\mathcal{D}$ given by
\begin{equation}\label{FW-Hamiltonian}
\mathcal{H} = \left ( \begin{matrix} \omega_{m} & \alpha  \\ \alpha  & \omega_{n} \\ \end{matrix} \right ) -i \left ( \begin{matrix} \gamma_{m} & \beta \\ \beta & \gamma_{n} \\ \end{matrix} \right ), \\
\end{equation} 	 	
\begin{equation}\label{FW-Scattering}
\mathcal{D} = \left ( \begin{matrix} d_{m1} & d_{n1}  \\ d_{m2}  & d_{n2} \\ \end{matrix} \right ) ,
\end{equation} 	 	
where $\alpha$ denotes the near-field coupling between the guided modes and $\beta$ represents the interference of radiating waves through far-field coupling. Matrix elements $d_{mj}$ and $d_{nj }$ represents the radiative coupling of $\mathrm{TE}_{m}$ and $\mathrm{TE}_{n}$ modes to the port $j$, respectively. Eigenmodes of the Hamiltonian are a linear combination of $\mathrm{TE}_{m}$ and $\mathrm{TE}_{n}$ modes, and from the determinant condition $|\mathcal{H} - \Omega \mathbf{I}|=0$, the corresponding eigenvalues are given by
\begin{equation}\label{FW-eigenvalue}
\Omega (k_z) = \bar{\Omega}(k_z) \pm\frac{1}{2} \sqrt{\left [ \Delta \Omega(k_z) \right ]^2 + 4(\alpha - i \beta)^2},
%= \bar{\Omega} \pm\frac{1}{2} \sqrt{x + iy},
\end{equation} 	
where $\bar{\Omega}=(\Omega_{m}+\Omega_{n})/2 $ and $\Delta \Omega = \Omega_{m}-\Omega_{n}$. From Eq.~(\ref{FW-eigenvalue}), we obtain avoided band structures in $k$ space.  Equation~(\ref{FW-eigenvalue}) indicates that the real parts of the two eigenvalues are degenerate, and the avoided band closes when the real part in the square root $x=(\Delta \omega)^2-(\Delta \gamma)^2 + 4(\alpha^2-\beta^2)$ is a negative value and the imaginary part $y=-2(\Delta\omega \cdot \Delta \gamma + 4\alpha \beta)$ is zero. When $\alpha =0$ with $0 < \rho < 1$, the band closes at $k_z=k_c$ because $y=0$ with $\Delta \omega (k_c)=0$ and $x=-(\Delta \gamma)^2-\beta^2$ is negative. In Fig.~\ref{fig2}(a) with $\rho_{0}=0.444$ and Fig.~\ref{fig3}(a) with $\rho_{0}=0.432$, two dispersion curves cross as straight lines at $k_c=k_d$ because near-field coupling vanishes with $\alpha=0$. For a given value of $\rho$, in the weakly modulated photonic lattice considered herein, the magnitudes of $\alpha$, $\beta$, and $\Delta \gamma = \gamma_{m} - \gamma_{n}$ are small and could be approximated as constant values near $k_c$, but $\Delta \omega = \omega_{m} - \omega_{n}$ changes from zero to some finite value as a function of $k_z$. When $\alpha \beta > 0$ is slightly deviated from zero with the variation of $\rho$ from $\rho_{0}$, the two conditions $y=0$ and $x<0$ can be fulfilled simultaneously at $k_z =k_d > k_c$ where $\Delta \omega \Delta\gamma < 0$, as shown in Figs.~\ref{fig2}(a) and ~\ref{fig3}(a) with $\rho=0.40$. When $\alpha \beta < 0$, on the other hand, bands can be closed at $k_z =k_d < k_c$ where $\Delta \omega \Delta\gamma > 0$ as shown in Figs.~\ref{fig2}(a) and ~\ref{fig3}(a) with $\rho=0.50$. The avoided band opens when the two conditions cannot be fulfilled simultaneously as $|\alpha \beta|$ is further increased with $0 < \rho < 1$.

\begin{figure}[]
\centering\includegraphics[width=10cm]{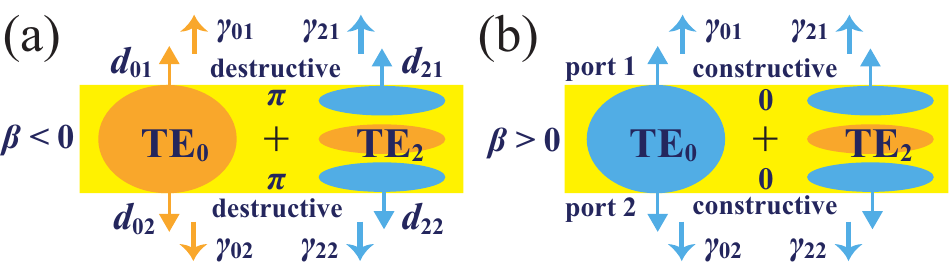}
\caption{Conceptual illustration of far-field coupling of radiating waves due to $\mathrm{TE}_{0}$ and $\mathrm{TE}_{2}$ modes. Radiating waves originating from different modes interfere (a) destructively when $\beta < 0$ and (b) constructively when $\beta > 0$ at the two radiation ports simultaneously.}
\label{fig4}
\end{figure}

Formation of the Friedrich-Wintgen BICs in Fig.~\ref{fig2}(c) and quasi-BICs in Fig.~\ref{fig3}(c) can be seen by determining $\beta$ in terms of decay rates. Due to the principle of energy conversation and time-reversal symmetry, the photonic structure shown in Fig.~\ref{fig1}(a) supports the relation $\mathcal{D}^\dagger \mathcal{D} = 2 \Gamma$, and by solving the relation, we have
%\begin{equation}\label{Condition-Gamma}
%\Gamma = \left ( \begin{matrix} \gamma_{m}  & \beta  \\ \beta  & \gamma_{n} \\ \end{matrix} \right ) = \left ( \begin{matrix} \gamma_{m1} + \gamma_{m2} & \beta  \\ \beta  & \gamma_{n1} + \gamma_{n2} \\ \end{matrix} \right )
%\end{equation} 	 	
%\begin{equation}\label{Condition-Scattering}
%\mathcal{C} \mathcal{D}^*=-\mathcal{D},
%\end{equation} 	
\begin{equation}\label{dm}
|d_{m1}|^2+|d_{m2}|^2=2\gamma_{m1} + 2\gamma_{m2},
\end{equation}
\begin{equation}\label{dn}
|d_{n1}|^2+|d_{n2}|^2=2\gamma_{n1} + 2\gamma_{n2},
\end{equation} 	 			 		
\begin{equation}\label{Beta-general}
|d_{n1}||d_{m1}|~e^{i(\theta_{n1}-\theta_{m1})} + |d_{n2}||d_{m2}|~e^{i(\theta_{n2}-\theta_{m2})}=2\beta,
\end{equation} 	 		
where $\theta_{mj}$ and $\theta_{nj}$ represent the phase angles of $d_{mj}$ and $d_{nj}$, respectively, and $\gamma_{mj}$ and $\gamma_{nj}$ denote the decay rates of $\mathrm{TE}_{m}$ and $\mathrm{TE}_{n}$ mode to the port $j$, respectively \cite{Kikkawa2019,WSuh2004}. Considering the avoided crossings between two even (even-like) modes shown in Fig.~\ref{fig2} (Fig.~\ref{fig3}), phase angles at port $1$ and port $2$ satisfy the relation $\exp (i\theta_{n1} - i\theta_{m1}) =\exp (i\theta_{n2} - i\theta_{m2}) = \pm 1$, as conceptually illustrated in  Fig.~\ref{fig4}. Moreover, it is reasonable to conjecture from Eqs.~(\ref{dm}) and (\ref{dn}) that $|d_{mj}|=\sqrt{2\gamma_{mj}}$ and $|d_{nj}|=\sqrt{2\gamma_{nj}}$. Hence, the far-field couplings between two even modes $\beta_{e-e}$ and between two even-like modes $\beta_{el-el}$ can be written as
\begin{equation}\label{Beta-even-even}
\beta_{e-e} = \pm \sqrt{\gamma_{n} \gamma_{m}},
\end{equation} 	 		
\begin{equation}\label{Beta-even-like-even-like}
\beta_{el-el} = \pm (\sqrt{\gamma_{n1} \gamma_{m1}} +  \sqrt{\gamma_{n2} \gamma_{m2}} ).
\end{equation} 	 		
In Eq.~(\ref{Beta-even-even}), we used $\gamma_{n1}=\gamma_{n2}=\gamma_{n}/2$ and $\gamma_{m1}=\gamma_{m2}=\gamma_{m}/2$. Coupled guided-mode resonance results in two hybrid eigenmodes. The anti-phase mode with $\beta < 0$ shown in Fig.~\ref{fig4}(a) can be a BIC or quasi-BIC because radiating waves from $\mathrm{TE}_{0}$ and $\mathrm{TE}_{2}$ modes interfere destructively at the two radiation ports simultaneously, and the in-phase mode with $\beta > 0$  in Fig.~\ref{fig4}(b) becomes more lossy because radiating waves interact constructively.

Maximal or minimal values of imaginary parts in the eigenvalues of the hybrid eigenmodes can be obtained when the two complex values $\Delta\Omega$ and $\alpha - i \beta$ in the square root of Eq.~(\ref{FW-eigenvalue}) are in phase, i.e.,
\begin{equation}\label{FW-condition}
\frac{ \Delta\gamma}{ \Delta\omega} = \frac{\beta}{\alpha}.
\end{equation}
With Eq.~(\ref{FW-condition}), Eq.~(\ref{FW-eigenvalue}) can be rewritten as
\begin{equation}\label{FW-my-eigenvalue-1}
%\Omega(k_z)=\bar{\Omega}(k_z) \pm \mu \sqrt{ (\alpha / \beta - i)^2},
\Omega(k_z)=\bar{\Omega}(k_z) \pm \mu (\alpha / \beta - i),
\end{equation}
where $\mu=\sqrt{(\Delta \gamma)^2 + 4 \beta^2}/2$ is a real positive value. In the photonic lattice with symmetric cladding layers, by Eq.~(\ref{Beta-even-even}), $\mu$ is the same as $-\mathrm{Im} (\bar{\Omega})=(\gamma_{m}+\gamma_{n})/2$, and the eigenvalue of anti-phase mode with $\beta < 0$ becomes purely real and turns into a BIC at $k_z=k_b=k_c=k_d$ when $\alpha =0$, as shown in Fig.~\ref{fig2}(b) with $\rho = \rho_{0}=0.444$. When $\alpha / \beta > 0$ ($\alpha / \beta < 0$), the Friedrich-Wintgen BICs with the anti-phase modes appear at $k_z = k_b < k_c < k_d$ ($k_z = k_b > k_c > k_d$) or at the lower (upper) band branch, as shown in Fig.~\ref{fig2}(b) with $\rho < \rho_{0}$ ($\rho > \rho_{0}$). In the photonic lattice with asymmetric cladding layers, by Eq.~(\ref{Beta-even-like-even-like}), $\mu$ is slightly different from $(\gamma_{m}+\gamma_{n})/2$. Therefore , when $\alpha =0$, a quasi BIC with the nonzero minimal imaginary part in the eigenfrequency appears at $k_z=k_b=k_c=k_d$, as shown in Fig.~\ref{fig3}(b) with $\rho = \rho_{0}=0.432$. When $\alpha / \beta> 0$ ($\alpha / \beta < 0$), the quasi BICs appear at $k_z = k_b < k_c < k_d$ ($k_z = k_b > k_c > k_d$) or at the lower (upper) band branch, as shown Fig.~\ref{fig3}(b) with $\rho < \rho_{0}$ ($\rho > \rho_{0}$).

\begin{figure}[]
\centering\includegraphics[width=10cm]{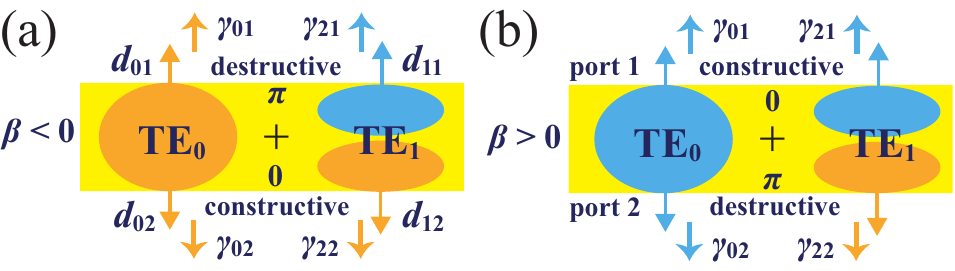}
\caption{Conceptual illustration of the far-field coupling of radiating waves due to $\mathrm{TE}_{0}$ and $\mathrm{TE}_{1}$ modes. (a) We set $\beta <0$ for convenience when radiating waves interact destructively (constructive) at the port 1 (port 2). (b) When $\beta > 0$, radiating waves interact constructively (destructive) at the port 1 (port 2). A coupled resonant mode could be an unidirectionally radiating mode whose decay rates to ports 1 and 2 is strongly asymmetric. }
\label{fig5}
\end{figure}

\begin{figure}[]
\centering\includegraphics[width=10.0cm]{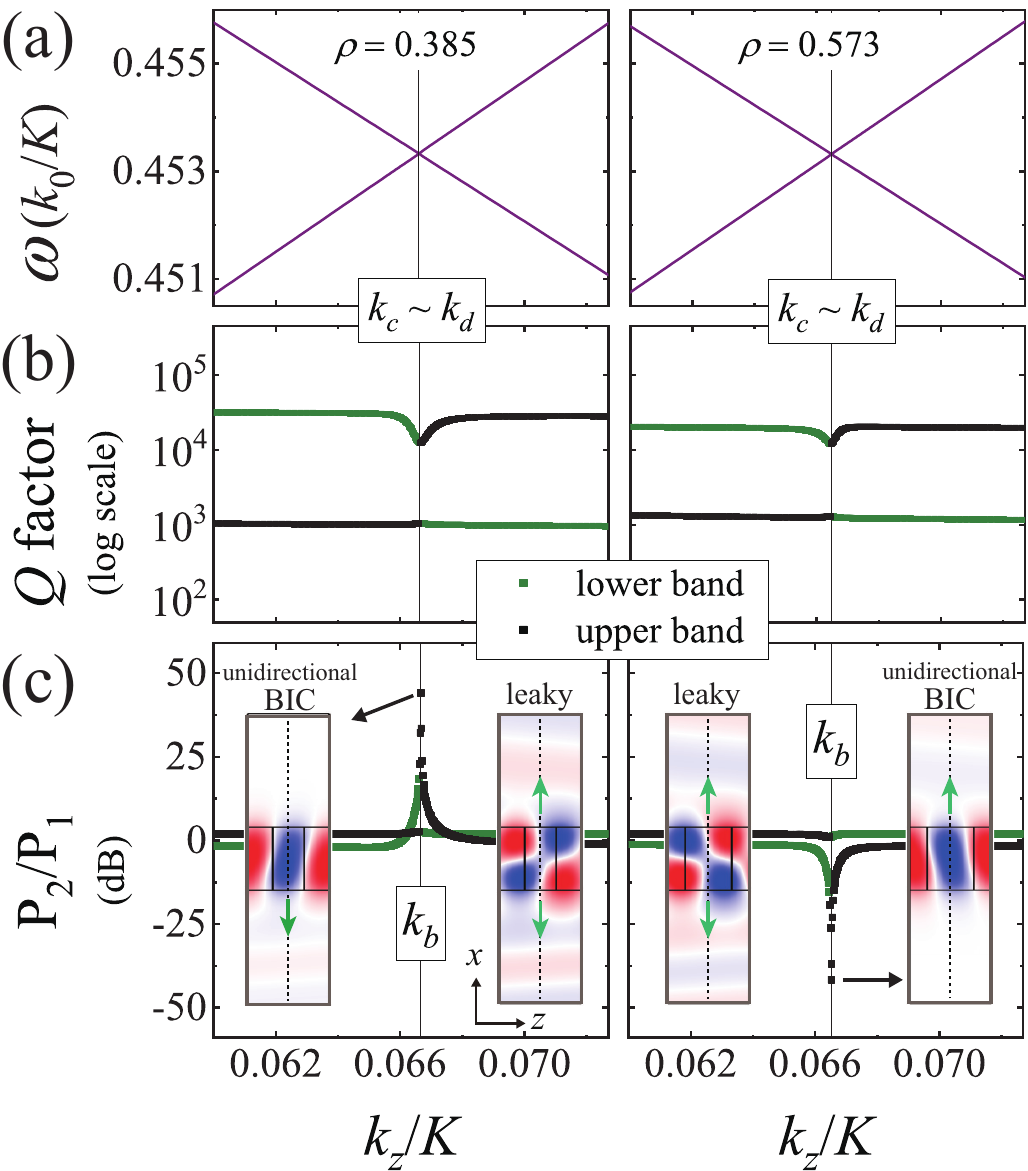}
\caption{ FEM simulated (a) band structures, (b) $Q$ factors, and (c) power ratios in leaky-mode photonic lattices with asymmetric cladding layers. Coupled guided-mode resonances result in hybrid eigenmodes composed of $\mathrm{TE}_{0}$ and $\mathrm{TE}_{1}$ modes near the crossing point $k_c$. Structural parameters are the same as in Fig.~\ref{fig3}. }
\label{fig6}
\end{figure}

When two guided modes with different transverse parities ($\mathrm{TE}_{0}$ and $\mathrm{TE}_{1}$) are coupling, as noted in Fig.~\ref{fig5}, radiating waves from different modes interfere constructively at one of the two radiation ports, while they interact destructively at the other port. Because Eqs.~(\ref{dm})--(\ref{Beta-general}) are valid for the coupling between two waveguide modes with different spatial parities, except that $\exp (i\theta_{n1} - i\theta_{m1}) = -\exp (i\theta_{n2} - i\theta_{m2}) = \pm 1$, the far-field coupling between an even and an odd mode and between an even-like and odd-like mode can be written as $\beta_{e-o} =0$ and
\begin{equation}\label{Beta-even-like-odd-like}
\beta_{el-ol} = \pm(\sqrt{\gamma_{n1} \gamma_{m1}} - \sqrt{\gamma_{n2} \gamma_{m2}}),
\end{equation} 	 		
respectively, where we set $\beta < 0$ ($\beta > 0$) when the radiating waves interfere destructively (constructively) at the port 1, for convenience. In the symmetric photonic lattices with $\beta_{e-o} =0$, near-field coupling $\alpha$ is also zero because the overlap integral of the even and odd modes is zero \cite{WSuh2004}. Two dispersion curves for the even and odd modes cross each other, and there is no band gap, as schematically represented in Fig.~\ref{fig1}(c). In photonic lattices with asymmetric cladding layers, on the other hand, avoided crossings due to $\mathrm{TE}_{0}$ and $\mathrm{TE}_{1}$ modes take place because $\alpha \neq 0$ and $\beta \neq 0$ in  general, and their properties can also be described by Eq.~(\ref{FW-eigenvalue}). Through FEM simulations, we verified  that a band gap opens at $k_c$, closes at $k_d$, closed band state remains for a while, reopens, and vanishes under variation of $\rho$ from 0 to 1. However, there cannot be a BIC or quasi-BIC due to the phase mismatch of the radiating waves at one of the two radiating ports, as shown in Fig.~\ref{fig5}. Instead, we found that there exists a unidirectional-BIC whose decay rate at one port is suppressed by the destructive interference, whereas decay to the opposite port is enhanced by constructive interaction. Figures~\ref{fig6}(a), \ref{fig6}(b), and \ref{fig6}(c) show the simulated band structures, $Q$ factors, and power ratios $\mathrm{P}_2/\mathrm{P}_1$, where $\mathrm{P}_j$ represents the radiation power to port $j$, respectively, when $\rho = 0.385$ and 0.583. Because the coupling strengths between even-like and odd-like modes are weak, as can be seen in  Fig.~\ref{fig6}(a), two dispersion curves cross as like straight lines at $k_d \sim k_c$ in the closed band states. Simulated $Q$ factors in Fig.~\ref{fig6}(b) show that there is no BIC or quasi-BIC. However, Fig.~\ref{fig6}(c) shows that there exist unidirectional-BICs whose radiation power to the port $1$ or port $2$ is significantly larger (up to 40 dB) than that to the opposite port. The spatial electric field distributions in the insets of Fig.~\ref{fig6}(c) demonstrate that unidirectional-BICs radiate to the only downward (upward) direction when $\rho =0.385$ ($\rho =0.573$), but leaky modes on the opposite band branches radiate to the upward and downward directions simultaneously. Here, we showed that unidirectional radiation can be enabled by unidirectional-BICs accompanied by avoided crossings. Very recently, unidirectional radiation has also been realized by utilizing the topological nature of BICs \cite{XYin2020}. We believe that the unidirectional radiation associated with BICs in planar photonic lattices is interesting and could be utilized to increase the efficiency of diverse optical devices, such as vertically emitting lasers and grating couplers.

\section{Conclusion}
In conclusion, we have investigated avoided crossings and BICs in 1D leaky-mode photonic lattices through FEM simulations and temporal coupled-mode theory. When two guided-mode resonances are coupled, photonic band gaps arise by avoided crossings and BICs appearing in photonic band structures without the fine tuning of structural parameters. The widths of avoided band gaps vary by lattice parameters. In particular, there exist closed band states in which avoided bands remain closed under variation of fill factor $\rho$. In photonic lattice slabs with symmetric cladding layers, true-BICs with, in principle, infinite $Q$ factor are accompanied by avoided crossings due to two guided modes with the same transverse parity. In the coupling process, two guided modes interact as in-phase or anti-phase. Anti-phase mode becomes a BIC because radiating waves from different modes vanish completely by destructive interference and in-phase mode gets more lossy with constructive interference. In photonic lattices with asymmetric cladding layers, on the other hand, only quasi-BICs with finite $Q$ factor are accompanied because the radiating waves by different modes cannot be completely eliminated. True- and quasi-BICs appear across the band gap by passing through a degenerate point where two dispersion curves cross as straight lines. We also show that unidirectional-BICs are accompanied by avoided crossings due to two guided modes with different transverse parities in asymmetric photonic lattices. The $Q$ factor of the unidirectional-BIC is finite but its radiation power in the upward or downward direction is significantly smaller than that in the opposite direction. Our research here is limited to the BICs and avoided crossings associated with the lowest three guided modes $\mathrm{TE}_{0}$, $\mathrm{TE}_{1}$, and $\mathrm{TE}_{2}$ in 1D photonic lattices. However, extension of this work to BICs and avoided crossings associated with higher order guided modes and 2D lattices is feasible. This contribution may be helpful in engineering BICs in diverse optical systems supporting leaky-modes.

This research was supported by a grant from the National Research Foundation of Korea funded by the Ministry of Education (No. 2020R1I1A1A01073945) and Ministry of Science and ICT (No. 2020R1F1A1050227), along with the Gwangju Institute of Science and Technology Research Institute in 2020.

\newpage

\end{document}